\documentclass[runningheads,a4paper]{llncs}

\usepackage{amssymb}
\usepackage{graphicx}
\usepackage{xcolor}
\usepackage{caption}
\usepackage{url}
\urldef{\mailsa}\path|{alfred.hofmann, ursula.barth, ingrid.haas, frank.holzwarth,|
\urldef{\mailsb}\path|anna.kramer, leonie.kunz, christine.reiss, nicole.sator,|
\urldef{\mailsc}\path|erika.siebert-cole, peter.strasser, lncs}@springer.com|    
\newcommand{\keywords}[1]{\par\addvspace\baselineskip
\noindent\keywordname\enspace\ignorespaces#1}

\usepackage{pifont}

\usepackage{float}

\usepackage{svg}
\usepackage{tikz}
\usepackage{pgf-pie}
\usepackage{pgfplots}
\usepackage{filecontents}

\usepackage{geometry}
\usepackage{tabularx}
\usepackage{booktabs}
\usepackage{multirow}

\begin{document}

\mainmatter  

\title{A systematic literature review of unsupervised learning algorithms for anomalous traffic detection based on flows}

\titlerunning{Review of unsupervised algorithms for anomalous traffic detection on flows}

%
%
\author{Alberto Miguel-Diez, Adrián Campazas-Vega, Claudia Álvarez-Aparicio, Gonzalo Esteban-Costales, Ángel Manuel Guerrero-Higueras}
\authorrunning{Alberto Miguel-Diez et al.}

\institute{Robotics Group. University of León, Campus de Vegazana S/N, 24071 León, Spain.
\\
\email{\{amigd, acamv, calvaa, gestc, agueh\}@unileon.es} \\
\url{https://robotica.unileon.es}}

%
%

\maketitle

\begin{abstract}

The constant increase of devices connected to the Internet, and therefore of cyber-attacks, makes it necessary to analyze network traffic in order to recognize malicious activity. Traditional packet-based analysis methods are insufficient because in large networks the amount of  traffic is so high that it is unfeasible to review all communications. For this reason, flows is a suitable approach for this situation, which in future 5G networks will have to be used, as the number of packets will increase dramatically. If this is also combined with unsupervised learning models, it can detect new threats for which it has not been trained. This paper presents a systematic review of the literature on unsupervised learning algorithms for detecting anomalies in network flows, following the PRISMA guideline. A total of 63 scientific articles have been reviewed, analyzing 13 of them in depth. The results obtained show that autoencoder is the most used option, followed by SVM, ALAD, or SOM. On the other hand, all the datasets used for anomaly detection have been collected, including some specialised in IoT or with real data collected from honeypots.

\end{abstract}

\section{Introduction}
\label{sec:introduction}

According to the Digital Report 2024 \cite{digital_report}, the number of Internet users in the world reached 5.35 billion people, a fairly high amount to which must be added those IoT or industrial devices that work autonomously making requests on the Internet. This increase in users also leads to an increase in cyber-attacks, according to Check Point this was 28\% in the first quarter of 2024 \cite{checkpointShiftingAttack}. Therefore, it is important to analyze network traffic in order to mitigate such threats, which are expected to be even greater with the arrival of 5G networks.

Traditionally, the method used to analyse network traffic was through the content of the packets that are transmitted. However, this method has two disadvantages: (i) it requires a large amount of computational load that is unfeasible with the arrival of 5G due to the high number of packets that will be transmitted in an instant of time and (ii) the solutions were based on checking if the packet complies with a pattern in order to mark it as anomalous. This meant that an updated list of all possible attack signatures had to be kept, which was difficult to maintain and prevented the detection of 0-day attacks. For this reason, flow-based analysis is used. A network flow is a collection of all packets sent between two hosts in a communication during an instant of time, so that their headers are then analysed to extract characteristics such as number of bytes sent, number of packets, protocol used, etc. Using flows, it is possible to determine malicious traffic generated by DoS attacks, botnets or malware.

Detection of anomalous can be done in different ways. In the early days of the development of these technologies, it was the network administrators who reviewed this traffic and categorised it as dangerous. However, as the traffic increased over time due to network improvements, this detection system became inefficient. Therefore, it started to detect anomalous traffic based on a set of rules predefined by the network administrators, if the analyzed traffic complies with these set of patterns, it is classified as anomalous \cite{Nisioti2018}. The latter has the disadvantage that it is not capable of detecting new threats that it has never seen before, as well as changing the rules so that they detect anomalous traffic based on what is unusual in this network.

Therefore, current solutions for detecting malicious traffic are based on machine learning techniques, supervised or unsupervised. The first one requires labelled datasets so that the models can learn the different network attacks. This data is difficult to gather because it requires a lot of human work to label the traffic, and it is also hard to find updated data. On the other hand, unsupervised models are trained with unlabeled data corresponding to normal network traffic. These models are able to detect any type of attack as an anomaly with respect to the normal pattern.  We have decided to use the latter approach in this study since it is more flexible than the supervised approach and is better adapted to the problem to be solved.


The fist step of the study was to look for existing reviews of the literature in this topic. Using exclusively network flows, the review by Sperotto, Schaffrath et al. (2010) \cite{Sperotto2010} details different flow-based intrusion detection systems divided by the type of attack they detect. Similarly, Sharma, Guleria, et al. (2018) \cite{Sharma2018} compile different methods for detecting malicious traffic based on flows, both machine learning and statistical methods. However, the review discusses both supervised and unsupervised learning methods.  Nisioti, Mylonas, et al. (2018) \cite{Nisioti2018} review discusses different systems to detect attacks that use unsupervised or hybrid learning to detect traffic, however, the solutions mixes both packet and flow-based traffic analysis. In addition, the datasets used for training are presented.  Yang, Liu et al. (2021) \cite{YANG2022102675} provide an analysis of different methods for intrusion detection of both supervised and unsupervised methods, choosing the articles based on the number of citations. 


Therefore, of the existing reviews in the literature, none of the reviews meet the research question to be addressed in this article. Furthermore, there are only two reviews that focus on flow-based anomaly detection, although none of them focus on unsupervised algorithms. These reviews are from 2010 and 2018 (\cite{Sperotto2010} and \cite{Sharma2018} respectively). In this work we have also analyzed 13 articles in depth that had not been analyzed in the previous reviews due to their publication date.


Therefore, the main contribution of this paper is a systematic review of the literature on unsupervised learning algorithms used to detect anomalous traffic in flows, as well as the datasets used to train them. This information is especially relevant to know which models work best for this task, as well as the most updated datasets to verify the accuracy of the solutions created. Thus, the following research questions (RQ) are proposed:

\begin{itemize}
    \item \textbf{RQ1} What are the unsupervised machine learning algorithms used for the detection of anomaly traffic from network flows?
    \item \textbf{RQ2} What datasets are used for training or testing the algorithms previously identified?
\end{itemize}

The work has been structured according to the PRISMA guidelines for the organisation of systematic reviews \cite{PRISMA}. Section \ref{sec:context} explains what network flows are and the advantages of using unsupervised learning to detect anomalies. Section \ref{sec:methodology} contains the protocol followed for the systematic review, while Section \ref{sec:results} presents the results of the research conducted. Based on the articles found, Section \ref{sec:discussion} answers the research questions posed above. Finally, the conclusions are found in Section \ref{sec:conclusions}.

\section{Context}
\label{sec:context}

The purpose of this section is to describe the main terms and concepts used in this work, formally presenting the concept of network flow, and different related protocols, as well as the use of unsupervised learning to detect anomalies, and its advantages to detect threats over rule-based, or legacy dataset-based systems.

\subsection{Flow Data}

A network flow depicts a set of packets passing through an observation point in the network during a specific time interval \cite{claise2004cisco}. All packets belonging to the same network flow share features such as source and destination address, source and destination port, and protocol type \cite{rfc7011}. Among the most commonly used lightweight flow-based protocols are NetFlow, SFlow, and IPFIX. They do not store packet payloads, so most of the information is lost in the communication. In return, the computational cost of analyzing it is reduced. 

NetFlow is a network protocol designed by Cisco to collect network traffic statistics \cite{dreijer2014}. It collects features such as the number of packets that make up a flow, the IP protocol type, protocol flags, and a timestamp. Many vendors also have their proprietary implementations of NetFlow, such as jFlow from Juniper and NetStream from Huawei. 

Sflow, stands for ``sampled flow'', is an industry standard for packet export at Layer 2 of the OSI model initially developed by InMon Corp~\cite{Reves2002}. It exports truncated packets and interface counters for network monitoring. The authoritative source of the sFlow protocol specifications is the sFlow.org consortium. sFlow takes a slightly different approach to network monitoring than other options. Where NetFlow statefully tracks flows, sFlow works by randomly sampling the full packet headers of a given flow at a predetermined interval. This feature can cut down on bandwidth and CPU utilization on the switch or router capturing flow information but may also reduce the accuracy of the information collected. sFlow does, however, capture deeper levels of information than NetFlow, including full packet headers and even partial packet payloads.  

IPFIX is an IETF protocol and the name of the IETF working group defines the protocol \cite{Hofstede2014}. 
It defines how IP flow information is formatted and transferred from an exporter to a collector. Previously, many data network operators relied on Cisco Systems' proprietary NetFlow technology for traffic flow information export. The IPFIX standards requirements were outlined in the original RFC 3917. Cisco NetFlow Version 9 was the basis for IPFIX. The basic specifications for IPFIX are documented in RFC 7011 through RFC 7015 and RFC 5103.

Table \ref{tab:comparison} summarizes the differences between the NetFlow, sFlow, and IPFIX protocols. As shown, all these protocols are open except for NetFlow. Another notable distinction is that while NetFlow and IPFIX are flow-based, sFlow is not. Additionally, all three protocols utilize sampling. NetFlow and IPFIX capture metadata and statistical information, whereas sFlow collects all packet headers and partial information from the packet contents.

\begin{table}
    \centering
    \resizebox{1\textwidth}{!}{
    \begin{tabular}{p{0.25\linewidth}p{0.25\linewidth}@{\hskip 0.2in}p{0.25\linewidth}@{\hskip 0.2in}p{0.25\linewidth}}
        \hline
         \textbf{Feature} & \textbf{NetFlow} & \textbf{sFlow} & \textbf{IPFIX} \\
         \hline
        Open or Proprietary & Proprietary & Open & Open \\
        \hline
         Flow Based & Yes & No & Yes \\
         \hline
         Sampled Based & Yes & Yes & Yes \\       
         \hline
         Information Captured & Metadata and statistical information, including bytes transferred, interface counters and so on (more details in \cite{cisco_netflow}) & Complete packet headers, partial packet payloads & Metadata and statistical information, including bytes transferred, interface counters and so on (more details in \cite{rfc7011})\\

         \hline
    \end{tabular}
    }
    \caption{Network flow protocols compared}
    \label{tab:comparison}
\end{table}



\subsection{Unsupervised machine learning in network flow data}


Machine learning techniques are being used to detect malicious traffic. On one hand, supervised learning techniques have been used with algorithms such as Stochastic Gradient Descent, Support Vector Machines, K-Nearest Neighbor, Gaussian Naive Bayes, Decision Tree, Random Forest, AdaBoost \cite{Igor2023}. These algorithms are trained on labeled data, learning patterns to classify new data into predefined categories. The problem with this  technique is that it is only capable of detecting known attacks, since it is trained with a dataset, which must contain all the types of attacks to be learned. Moreover, finding up-to-date labelled datasets is difficult, so these models are often trained on outdated data.

The use of unsupervised learning has started to gain relevance when it comes to detecting anomalies. It is a machine learning technique where the algorithm learns patterns from unlabeled data without explicit guidance. Instead of being given correct outputs, the algorithm explores the structure of the data to find hidden patterns or relationships. In the context of detecting anomalies, unsupervised learning involves identifying deviations or outliers from the norm within the dataset without prior knowledge of what constitutes normal behavior.

The advantages of this approach are described in the following points \cite{Nisioti2018}:

\begin{itemize}
    \item Unsupervised systems are able to detect unknown attacks, e.g. 0-day attacks. Remark that supervised systems are not good at this task. 
    \item Unsupervised methods spare the time-consuming training regimen required by supervised methods, offering a more efficient approach.
    \item Unsupervised methods do not require labelled data, which are difficult and resource-consuming to obtain. In addition, datasets that are labelled have old attacks, which can make it difficult to detect current attacks.
\end{itemize}

\section{Methodology}
\label{sec:methodology}

PRISMA \cite{PRISMA} methodology and the recommendations by Kitchenham \cite{Kitchenham} have been used to carry out this systematic review. The PRISMA statement was published to set standards for reporting systematic reviews and meta-analyses. This statement was designed to help authors of systematic reviews to transparently document why the review was conducted, what the authors did, and what they found. In addition, Kitchenham proposes guidelines for performing systematic literature reviews in software engineering to guide researchers in evaluating and interpreting all available research literature for a particular research question. Kitchenham's recommendations have three stages:

\begin{enumerate}
    \item Planning, where the research question is posed, and the review protocol is set.
    \item Review, where relevant studies are identified and their quality is assessed -- this phase ends with the extraction of data from the articles studied --.
    \item Documentation, where a report of the review is written.
\end{enumerate}

The above stages have been divided into the following subsections: search planning (1), search process (2), sample selection (3), and data extraction (4).

\subsection{Search planning}

The literature reviews presented in Section \ref{sec:introduction} have been read and the research questions they answer have been extracted. Subsequently, they have been compared with those to be answered in this article and it has been seen that they do not coincide, which highlights the new contributions provided by this literature review.

Next, the databases containing the articles that will comprise the literature review should be chosen. It was decided to use IEEE Digital Library, Scopus, and Web of Science. This is because they are reputable sources that contain peer-reviewed articles and are widely used in the field of computer science.

\subsection{Search process}

The search process began by taking as a baseline a set of semi-reference articles (see Table \ref{tab:articulos_semi_ref}). This set is based on the proposal of Zhang, Babar and Tell \cite{ZHANG2011625}; although in this case, they were obtained informally before the systematic literature review process. Based on the previously described articles, an automatic search was planned until April 2024 in the electronic article databases described before. In all of these, the search was filtered by articles belonging to a journal or conference written in English or Spanish. In addition, these searches were based on the appearance of certain terms in the title, abstract, and keyword sections.

\begin{table}[htb]
\centering
\caption{Group of semi-reference articles}
\label{tab:articulos_semi_ref}
\resizebox{1\textwidth}{!}{
\begin{tabular}{c l }
\hline
 \textbf{Reference} & \textbf{Title} \\
\hline
 \cite{kabir_semiref} & Unsupervised Learning for Network Flow Based Anomaly Detection in the Era of Deep Learning \\
 \cite{Malaiya_semiref} & An Empirical Evaluation of Deep Learning for Network Anomaly Detection \\
 \cite{truonghuu_semiref} & An Empirical Study on Unsupervised Network Anomaly Detection using Generative Adversarial Networks \\
 \cite{Verkerken_semiref} & Unsupervised Machine Learning Techniques for Network Intrusion Detection on Modern Data \\
\hline
\end{tabular}
}
\end{table}

Then, the chain was applied in the databases defined above. It was built using the keywords extracted after applying the PICOC strategy \cite{picoc1,picoc2} to the research question posed in Section \ref{sec:introduction}. As its name suggests, PICOC is a method used to describe the five elements of a searchable question: (1) population, (2) intervention, (3) comparison, (4) outcomes, and (5) context.

The search string selected to carry out our research is as follows:

\begin{quote}\it
    ( `anomaly detection' \textbf{OR} `malicious traffic' ) \textbf{AND} ( `machine learning' \textbf{OR} `deep learning' ) \textbf{AND} unsupervised \textbf{AND} ( `NetFlow' \textbf{OR} `IPFIX' \textbf{OR} `sFlow' \textbf{OR} `flow-based' ) \textbf{AND} ( method \textbf{OR} technique \textbf{OR} procedure \textbf{OR} algorithm )

\end{quote}

Finally, the automatic search was complemented with a process of snowball sampling, which consisted of verifying the references of those articles reviewed after the selection process and that directly answered the research questions.

\subsection{Selection of samples}

In order to filter the results previously obtained with the search string, it was decided to carry out a selection process based on a set of inclusion and exclusion criteria. If an article meets an exclusion criterion and an inclusion criterion, the exclusion criterion will be given more weight.

In the first phase, this selection process was based on reading the title, abstract and keywords. In some cases, this is not enough to know if the article answers the research questions, for this reason, after eliminating duplicates, a second phase was carried out in which copies of the rest of the articles were obtained in order to completely review their text.

The Inclusion Criteria (IC) are depicted below. Articles that do not meet any of the inclusion criteria are removed.

\begin{itemize}
    \item \textbf{IC1} The article uses machine learning with network flows for the detection of anomaly traffic.
\end{itemize}

Finally, the  Exclusion Criteria (EC) are listed below. Articles that meet at least one exclusion criterion are removed.

\begin{itemize}
    \item \textbf{EC1} The article does not specify the use of supervised or unsupervised learning.
    \item \textbf{EC2} The article does not use network flows.
    \item \textbf{EC3} The article uses a semi-supervised machine learning algorithm.
    \item \textbf{EC4} The article uses machine learning based on flows but not on the investigated topic.
    \item \textbf{EC5} The article uses supervised machine learning.
    \item \textbf{EC6} The results presented in the article are the same as those published by the same authors.
\end{itemize}

Given the substantial volume of articles meeting the criteria outlined above, it was deemed necessary to establish a quality assessment process to identify those that best address the research questions. This involved assigning a score ranging from 0 to 5 points to each article, with a minimum acceptance threshold of 3.5 points. Consequently, articles falling below this threshold would be excluded. The questionnaire consisted of the following set of questions (Q):

\begin{itemize}
    \item \textbf{Q1} Is only flow data used to predict anomalous traffic?
    \item \textbf{Q2} Are the names of the machine learning algorithms used mentioned?
    \item \textbf{Q3} Are the datasets they work with specified?
    \item \textbf{Q4} Is the algorithm tested with a dataset known in the literature?
    \item \textbf{Q5} Are various algorithm metrics provided in tabular form?
\end{itemize}

The questions posed above allow for selecting the most relevant articles for the research carried out in this work. Selected papers will allow us to learn about the most widely used unsupervised machine learning algorithms and the ones that provide the best results in detecting malicious network traffic on network flow data. Besides, we will pinpoint the most commonly used datasets and KPIs to measure the efficiency of the solution proposed in the papers.

There are two questions that are relevant when finding the datasets used. Question Q3 seeks to reward those articles that indicate the dataset used, while question Q4 aims to find that the dataset used can be downloaded and/or is widely used in the literature. The latter question seeks to discard those articles that have created their dataset, and do not share the data which does not allow to reproduce the study.

Once the questionnaire is ready, its evaluation criterion is established. We grant a one-point value per Yes-answered question and a zero-point value for No-answered questions. A partial answer may also be given with a score of half a point. Therefore, an article can get a maximum of five points. 

\subsection{Data extraction}

Once the study universe has been narrowed down, a data extraction form is established. The features to be extracted from each paper come from the research question posed in Section \ref{sec:introduction}. The data is extracted from each paper after a complete and detailed reading. 
Thus, the features (F) to be extracted are as follows:

\begin{itemize}
    \item \textbf{F1} Name of the software solution (if available)
    \item \textbf{F2} What type of anomalies does the algorithm detect?
    \item \textbf{F3} Machine learning algorithms used in the article
    \item \textbf{F4} KPI of the algorithm(s) used in the research
    \item \textbf{F5} Features
    \item \textbf{F6} Name of the dataset(s) used for testing
    \item \textbf{F7} Year of the dataset(s)
\end{itemize}

\section{Results}
\label{sec:results}

\begin{figure}[htb]
	\centering{
		\includegraphics[width=3.0in]{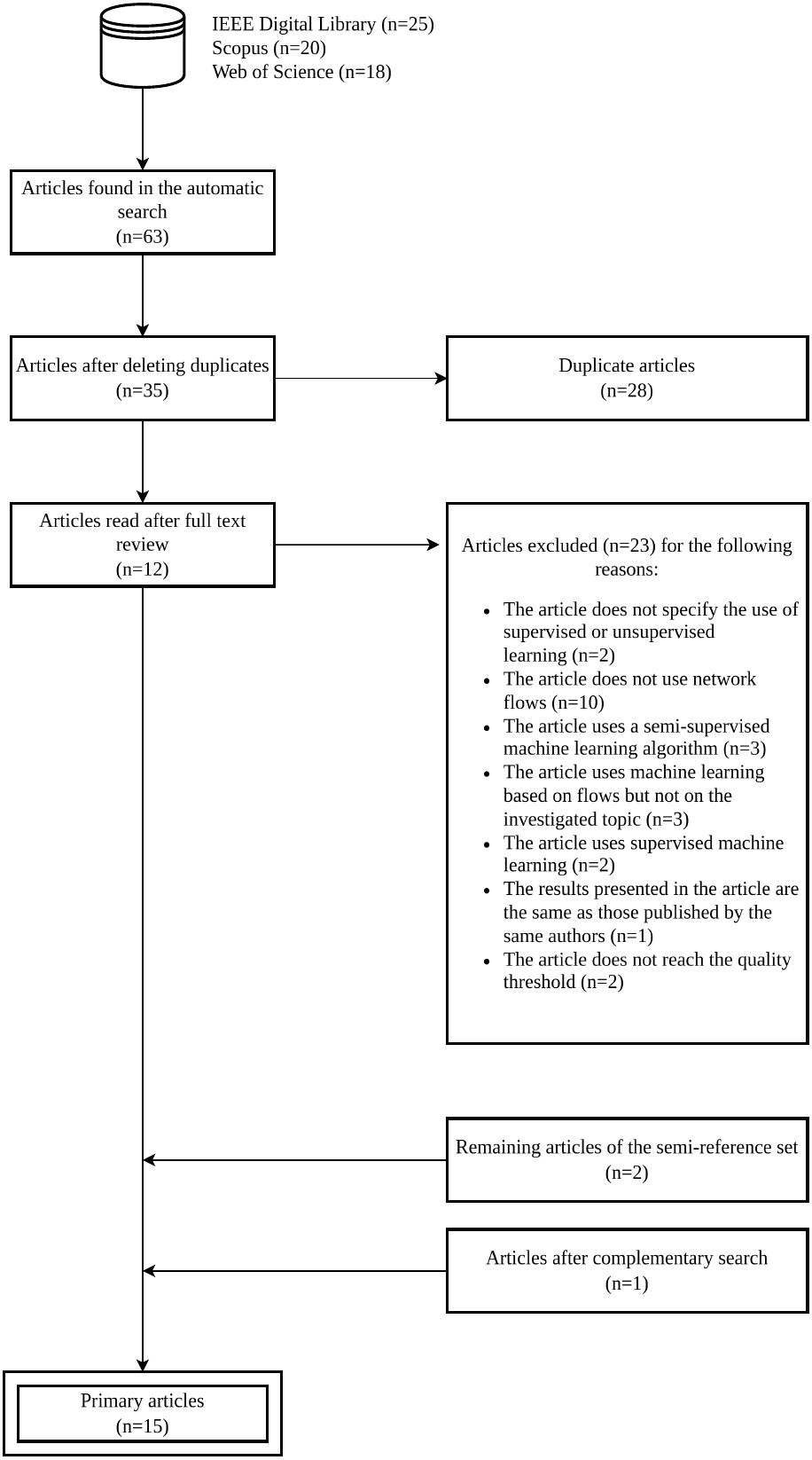}
	}
	\caption{Diagrama PRISMA}
	\label{fig:diagrama_prisma}
\end{figure}

The result of the process described in  Section \ref{sec:methodology} can be seen in the PRISMA diagram in Figure \ref{fig:diagrama_prisma}.

The first stage consists of selecting the sources of information. The sources used in the research are IEEE Digital Library, Scopus and Web of Science. Next, we had to carry out the search process by applying the search string in the above sources of information. 

In this first automatic search phase, a total of 63 articles were found.  Afterwards, in the selection phase, 28 duplicate articles were discarded. Therefore, the remaining 35 articles were carefully reviewed by reading their content. From these, 21 articles were rejected because they met some exclusion criterion 
and another 2 articles because they did not exceed the imposed quality threshold.



Therefore, the set of main articles consisted of 12 papers, to which was added the set of semi-reference articles consisting of 2 additional papers. This set, which was obtained manually at the beginning, consisted of 4 articles, 2 of which were duplicates because they were also obtained with the automatic search. Therefore, 50\% of the semi-reference articles were found with the automatic search.

Finally, a complementary search was performed on the references of the remaining 14 articles. During this procedure, one more relevant article was found to add to the review. Thus, a total of 15 primary articles were identified at the end of the entire review process.

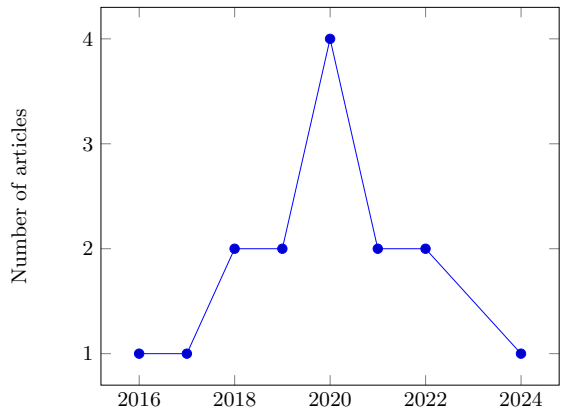
\begin{figure}[htb]
	\centering{
            \resizebox{0.5\linewidth}{!}{
            
		\begin{tikzpicture}
            \begin{axis}[
            ylabel=Number of articles,
            legend pos=north east,
            xticklabel style={/pgf/number format/1000 sep={}
            }
            ]
              \addplot table [x=Year,y=Number] {data.txt};
            \end{axis}
            \end{tikzpicture}
            }
	}
	\caption{Trend graphs of selected articles}
	\label{fig:diagrama_anyos}
\end{figure}

In Figure \ref{fig:diagrama_anyos} it can be seen that the selected articles were published between the years 2016 and 2024. In addition, there is a growth of articles in the year 2020 with four publications, while in the year 2023, there are no articles selected.

Regarding the evaluation of the quality of the results, Table \ref{tab:quality_articles} compiles the 15 primary articles found. The lowest score recorded is 4 points, while the highest score is 5 points, which was obtained by 9 articles. It should be noted that the two articles that did not exceed the quality threshold (\cite{Jadidi2022} and \cite{Zhang2017}) obtained 3 and 2 points respectively.

\begin{table}[htb]
\centering
\begin{tabular}{c c c c c c c}
\hline
 \textbf{Reference} & \textbf{Q1} & \textbf{Q2} & \textbf{Q3} & \textbf{Q4} & \textbf{Q5} & \textbf{Total}\\
\hline
    \cite{Chen2017} & 1 & 1 & 1 & 1 & 1 & 5\\
    \cite{GustavoGonz2021} & 1 & 1 & 1 & 1 & 1 & 5\\
    \cite{Gorbett2022} & 1 & 1 & 1 & 1 & 1 & 5\\
    \cite{kabir_semiref} & 1 & 1 & 1 & 1 & 1 & 5\\
    \cite{Kompougias2021} & 1 & 1 & 1 & 1 & 1 & 5\\
    \cite{Li2024} & 1 & 1 & 1 & 1 & 1 & 5\\
    \cite{Malaiya_semiref} & 1 & 1 & 1 & 1 & 1 & 5\\
    \cite{truonghuu_semiref} & 1 & 1 & 1 & 1 & 1 & 5\\
    \cite{Verkerken_semiref} & 1 & 1 & 1 & 1 & 1 & 5\\

    \cite{GarciaCordero2016} & 1 & 1 & 1 & 1 & 0 & 4\\
    \cite{Kotani2018} & 1 & 1 & 1 & 1 & 0 & 4\\ 
    \cite{Nguyen2019} & 1 & 1 & 1 & 1 & 0 & 4\\
    \cite{Schueller2018} & 0 & 1 & 1 & 1 & 1 & 4\\
    \cite{Timcenko2022} & 1 & 1 & 1 & 1 & 0 & 4\\
    \cite{Wei2020} & 1 & 1 & 1 & 0 & 1 & 4\\
    \cite{Jadidi2022} & 1 & 1 & 0 & 0 & 1 & 3\\
    \cite{Zhang2017} & 1 & 1 & 0 & 0 & 0 & 2\\
\hline
\end{tabular}
\caption{Results of the quality of primary and discarded articles}
\label{tab:quality_articles}
\end{table}

\section{Discussion}
\label{sec:discussion}

\subsection{Research question RQ1}

The primary aim of research question RQ1 outlined in Section \ref{sec:introduction} is to identify unsupervised learning algorithms employed for the detection of anomalous network traffic, of any type, within network flows. It is pertinent to underscore that the scope of machine learning algorithms considered encompasses deep learning algorithms, a subset within this domain. The solutions presented are not hybrid, that is, they do not use supervised and unsupervised learning simultaneously.  This deliberate choice is motivated by the pursuit of adaptable solutions independent of human oversight, capable of discerning novel threats that may not have been previously documented.

The solution proposed by \cite{kabir_semiref} performs a comparative analysis using several machine learning algorithms: k-means, self-organizing maps (SOM), deep autoencoding gaussian mixture model (DAGMM) and adversarially learned anomaly detection (ALAD). Analyzing the results presented by the authors, it is observed that DAGMM has the lowest false positive rate (0.9\%) and a relatively high detection rate of 99.0\% in the KDD dataset. The tests were also performed with the ISCX dataset showing that SOM obtains the best results with a detection rate of 95.6\% and a false positive rate of 7.9\%. Finally, k-means, although not obtaining the best results, was able to detect attacks for which there was no instance in the training set.

In \cite{Kotani2018} an algorithm is proposed to detect intrusions using robust autoencoders (RAE) that decrease false positives in the test data by removing outliers in the training data. However, although it is observed that the FPR is reduced and attacks are more easily detected, no objective metric is provided to determine this.

The authors of \cite{Timcenko2022} propose a system to detect anomalous traffic based on two models, entropy-based pre-processing and expectation-maximization algorithm. The last is a soft clustering based on the calculation of the probability density to more accurately allocate data to a specific cluster, without strict limitations between the clusters. The chosen characteristics of the flows are obtained based on the entropy calculations. In addition, an architecture is proposed to implement the created model in an enterprise network.

In \cite{Schueller2018}, a model based on the Support Vector Machine (SVM) is proposed to detect any type of anomalies. This is a hybrid system in which both flows and packets are analyzed. First, flows are analyzed and when an attack is detected, the packets are analyzed. Therefore, flow analysis is the first line of defence. This reduces the number of false positives. The detection rate using the flow-based IDS is 84.78\% and using the combination of flows and packets is 83.81\%. The authors used a testbed created with Mininet network simulator \cite{mininet} consisting of five switches and 16 servers plus a dataset.

The authors of the paper \cite{Kompougias2021} present a system to detect IoT botnets. The framework is based on two autoencoders: (i) a vanilla implementation of a deep autoencoder and (ii) GANomaly, a combination of autoencoders with conditional generative adversarial networks (GANs). With the second one and a decision boundary of 0.001\%, the precision obtained is 92\% and with autoencoder it is 75\%.

In the solution proposed in \cite{Chen2017}, botnet traffic is detected using self-organizing maps, local outliers and k-NN outlier factors. Using these systems to detect botnets is interesting because botnets use encrypted payloads and change ports dynamically in order to pass signature detection and packet analysis. Regarding the training of the algorithm, only benign traffic is used to know the normal operation of the system. The detection rate is 91\% with a false positive rate of 5\%.

In \cite{GarciaCordero2016}, a model using an extended version of the Replicator Neural Network (RNN) is proposed, this type of network is used as a compression technique. This process is related to the PCA dimensionality reduction technique \cite{HechtNielsen1995}, which is useful in flows due to the high number of features obtained. The authors, however, do not provide any numerical metric to know the efficiency of the solution, they only provide graphs (although it can be observed that the results obtained are favourable).

In \cite{GustavoGonz2021}, I-LADS  is a proposed solution for anomaly detection using AutoEncoders (AE). It is available in two versions: (i) I-LADS-v1, which employs filters to individually model IP addresses within the NetFlow dataset, enabling the training of a distinct model for each filtered IP address; and (ii) I-LADS-v2, which operates without filters, thus training a single algorithm for all IP addresses. The authors compare the same models using supervised and unsupervised learning, concluding that V1 works better with supervised and V2 better with unsupervised. The accuracy obtained with I-LADS-v2 is 0.9371 and I-LADS-V1 is 0.7742.

The solution proposed in \cite{Verkerken_semiref} uses four different types of machine learning algorithms: principal components analysis, isolation-forest, one-class support vector machine and autoencoder. The highest precision is obtained by the isolation-forest algorithm (0.9470), while the highest AUROC is obtained by the autoencoder method (0.978). Furthermore, according to the authors, all four approaches have acceptable computational complexities to be used in real-world applications with training and inference time in the order of seconds to minutes. On the other hand, it is important to highlight, that all proposed methods were able to detect the ``heartbleed attacks'' which has only 11 occurrences in the CIC-IDS-2017 dataset.

In \cite{Nguyen2019}, a model is proposed to detect and explain anomalies in network traffic. A Variational Autoencoder (VAE) is used to detect anomalies and a gradient-based fingerprinting technique is used to explain anomalies. This is achieved by examining the gradients ``provided'' by each feature of the data point, which are obtained from the VAE by automatic differentiation in TensorFlow. The ROC metric for VAE to detect various DoS attacks, BotNet, Scan11, Scan44, and Spam has a mean of 0.947.

FlowGANAnomaly \cite{Li2024} allows to detect anomalies with a generator and a discriminator. Instead of simply detecting the anomalies by the output of D, they proposed a new anomaly scoring method that integrates the deviation between the output of two Gs’ convolutional encoders with the output of D as weighted scores to improve the low recall rate of anomaly detection. They obtain an accuracy of 0.9888 with the dataset NSL-KDD.

The authors of \cite{Gorbett2022} are able to detect general IoT anomalies using a Weighted Hamming Distance LID Estimator, an algorithm created by the authors derived from Local Intrinsic Dimensionality (LID). Their model was compared with Isolation Forest, KNN, Weighted Hamming KNN and Autoencoder and in all except in the precision-recall curve with the ToN-IoT dataset they obtained better results. In that case, Autoencoder had a PR of 0.985. For example, with the NF-BoT-IoT dataset, a ROC value of 0.970 and a PR of 0.999 is achieved.

In \cite{truonghuu_semiref}, generative adversarial networks (GANs) are used adopting two architectures: AnoGAN and ALAD.  It has been shown in the paper that the latter architecture performs the best with a precision of 0.8473 on the UNSW-NB15 dataset and an AUROC of 0.9882. In addition, it follows a new feature concerning other frameworks, as well as aggregating the traffic in flows, it aggregates it in sessions. A traffic session comprises flows sharing identical 5-tuple attributes (source IP, destination IP, source port, destination port, protocol), with the condition that the time gap between consecutive flows is below a specified threshold.

In the article \cite{Malaiya_semiref}, a comparison of the efficiency of the following models is made: Fully Connected Network (FCN), Variational AutoEncoder (VAE) and Long Short-Term Memory with Sequence to Sequence (LSTM Seq2Seq). An accuracy of 99\% has been achieved using an LSTM Seq2Seq structure.

Finally, BotMark \cite{Wei2020} is a proposed solution to detect botnets using hybrid analysis: flow-based and graph-based network traffic behaviours. The authors decided to apply preprocessing by eliminating those flows that are not correctly established, i.e. those flows that do not complete the TCP hand-shake and those that are not from internal to external hosts. As for the metrics, an accuracy of 0.9994 is achieved. 

A summary of the answer to research question RQ1 is shown in Table \ref{tab:algorithms_comparison} in which all the unsupervised machine learning algorithms used to detect anomalies in network flows are grouped.

\begin{table}[htb]
    \centering
    \resizebox{\textwidth}{!}{
    \begin{tabular}{c c  c  c  c  c  c  c c c}
        \hline
       \textbf{Algorithm}  & \textbf{Reference} & \textbf{Dataset} & \textbf{Accuracy} & $\mathcal{P}$ & $\mathcal{R}$ & $\mathcal{F}_1$ & AUROC & DR & FPR\\
       \hline
       \multirow{2}{*}{K-Means} & \multirow{2}{*}{\cite{kabir_semiref}} & KDD-CUP'99 & -- & -- & -- & --& -- & 0.998 & 0.031 \\
       & & UNB ISCX IDS & -- &  --& --& --& --&  0.953& 0.097\\
       \hline
       \multirow{3}{*}{SOM} & \multirow{2}{*}{\cite{kabir_semiref}} & KDD-CUP'99 & -- & -- & -- & --& -- & 0.999 & 0.09 \\
       & & UNB ISCX IDS & -- &  --& --& --& --&  0.956& 0.079\\
       \cline{2-10}
       & \cite{Chen2017}& CTU-13 & -- &  --& --& --& --&  0.732& 0.051\\
       \hline
       \multirow{2}{*}{DAGMM} & \multirow{2}{*}{\cite{kabir_semiref}} & KDD-CUP'99 & -- & -- & -- & --& -- & 0.99 & 0.09 \\
       & & UNB ISCX IDS & -- &  --& --& --& --&  0.943 & 0.358\\
       \hline
       \multirow{5}{*}{ALAD} & \multirow{2}{*}{\cite{kabir_semiref}} & KDD-CUP'99 & -- & -- & -- & --& -- & 0.995 & 0.271 \\
       & & UNB ISCX IDS & -- &  --& --& --& --&  0.863 & 0.547\\
       \cline{2-10}
       & \multirow{3}{*}{\cite{truonghuu_semiref}} & UNSW-NB15 & -- &  0.8473 & 0.8583 & 0.8527 &  0.9882 & -- & -- \\
       &  & CIC-IDS-2017 & -- &  0.8260 & 0.8268 & 0.8264  &  0.9529 & -- & -- \\
       &  & Stratosphere IPS  & -- &  0.6993 & 0.6996 & 0.6995  &  0.9053 & -- & -- \\
       \hline
       \multirow{6}{*}{Autoencoders} & \cite{Kotani2018} & MAWI & -- & -- & -- & --& -- & -- & --\\
       \cline{2-10}
        & \cite{Kompougias2021} & CIC-IDS-2017 & -- & 0.75 & -- & --& -- & -- & --\\
       \cline{2-10}
        & \cite{Verkerken_semiref} & CIC-IDS-2017 & -- & 0.9459 & 0.9778 &  0.9616 & 0.9775 & -- & --\\
       \cline{2-10}
        & \cite{GustavoGonz2021} & CIDDS-001 & 0.9371 & 1.0000 & 0.9274 &  0.9623 & -- & -- & --\\
       \cline{2-10}
        & \cite{Nguyen2019} & UGR16 & -- & -- & -- & --& 0.947 & -- & --\\
       \cline{2-10}
        & \cite{Malaiya_semiref} & Kyoto University & -- & 0.981 & 0.901 & 0.939 & -- & -- & --\\
       \hline
       Entropy-based pre-processing & \multirow{2}{*}{\cite{Timcenko2022}} & \multirow{2}{*}{CTU-13} & \multirow{2}{*}{--} & \multirow{2}{*}{--} & \multirow{2}{*}{--} & \multirow{2}{*}{--} & \multirow{2}{*}{--} & \multirow{2}{*}{--} & \multirow{2}{*}{--} \\
       and EM algorithm & & \\
       
       \hline
       \multirow{2}{*}{SVM} & \cite{Schueller2018} & DARPA & -- & -- & -- & --& -- & 0.8478 & --\\
       \cline{2-10}
        & \cite{Verkerken_semiref} & CIC-IDS-2017 & -- & 0.9104 & 0.9920 & 0.9495 & 0.9705 & -- & --\\
       \hline
       
       GANomaly & \cite{Kompougias2021} & CIC-IDS-2017 & -- & 0.92 & -- & --& -- & -- & --\\
       \hline
       LOF & \cite{Chen2017} & CTU-13 & -- & -- & -- & --& -- & 0.383 & 0.053\\
       \hline
       k-NN & \cite{Chen2017} & CTU-13 & -- & -- & -- & --& -- & 0.913 & 0.051\\
       \hline
        Replicator Neural Network & \cite{GarciaCordero2016} & MAWI & -- & -- & -- & --& -- & -- & --\\
       \hline
        Principal Components Analysis & \cite{Verkerken_semiref} & CIC-IDS-2017 & -- & 0.9346 & 0.9435 & 0.9390 & 0.9373 & -- & --\\
       \hline
        Isolation Forest & \cite{Verkerken_semiref} & CIC-IDS-2017 & -- & 0.9470 & 0.9314 & 0.9391 & 0.9584 & -- & --\\
       \hline
        \multirow{4}{*}{FlowGANAnomaly} & \multirow{4}{*}{\cite{Li2024}} & CIC-IDS-2017 & 0.8840 & 0.9841 & 0.8840 & 0.9295 & 0.7432 & -- & --\\
         &  & NSL-KDD & 0.8747 & 0.9888 & 0.8747 & 0.9245 & 0.9801 & -- & --\\
         &  & UNSW-NB15 & 0.7354 & 0.9859 & 0.7037 & 0.8384 & 0.8530 & -- & --\\
         &  & CIC-DDoS2019 & 0.9012 & 0.8590 & 0.9012 & 0.8727 & 0.7883 & -- & --\\
        \hline
       \multirow{3}{*}{Weighted Hamming Distance LID} & \multirow{3}{*}{\cite{Gorbett2022}} & NF-BoT-IoT & -- & -- & -- & -- & 0.970 & -- & --\\
        &  & ToN IoT & -- & -- & -- & -- & 0.985 & -- & --\\
        &  & Aposemat IoT-23  & -- & -- & -- & -- & 0.998 & -- & --\\
       \hline

       \multirow{3}{*}{AnoGAN} & \multirow{3}{*}{\cite{truonghuu_semiref}} & UNSW-NB15 & -- & 0.4345 &  0.4394 & 0.4369 & 0.8765 & -- & --\\
        &  & CIC-IDS-2017 & -- & 0.1439  & 0.1440 &  0.1439 & 0.7492 & -- & --\\
        &  & Stratosphere IPS & -- & 0.4676 &  0.4678 &  0.4677 &  0.8493 & -- & --\\
       \hline

        Fully Connected Network & \cite{Malaiya_semiref} & Kyoto University & -- & 0.997 &  0.874 & 0.931 & -- & -- & --\\
        \hline

        LSTM & \cite{Malaiya_semiref} & Kyoto University & -- & 1 &  1 & 1 & -- & -- & --\\
        \hline

        Flow-based \& graph-based & \cite{Wei2020} & Manual & 0.9994 & -- &  -- & 0.115207 & -- & -- & --\\
        \hline
    \end{tabular}
    }
    \caption{Algorithms Comparison. The following KPIs are shown: Accuracy, precision ($\mathcal{P}$), recall ($\mathcal{R}$), F$_1$ score ($\mathcal{F}_1$), AUROC, detection rate (DR) and false positive rate (FPR)}
    \label{tab:algorithms_comparison}
\end{table}

\subsection{Research question RQ2}

Regarding the datasets, Table \ref{tab:datasets_discussion} shows the features of the datasets. Specifically, each dataset indicates the creation date, whether or not the dataset is open or proprietary, the paper in which it is referenced, and whether the content is real or synthetic. As shown in the table, 16 datasets were identified. CIC-IDS-2017 is the most popular since it is mentioned in four papers. 

The table shows that most datasets contain flow data from synthetic network traffic. It is usual since, otherwise, network flow data could not be empirically labelled as benign or malicious. 

\begin{table}[htb]
\centering
\resizebox{1\textwidth}{!}{
\begin{tabular}{c@{\hskip 0.3in} c@{\hskip 0.3in} c@{\hskip 0.3in} c@{\hskip 0.3in}  c@{\hskip 0.3in}}
\hline
 \textbf{Dataset} & \textbf{Date} & \textbf{Open/Proprietary} & \textbf{Reference/s } & \textbf{Real/Synthetic}\\
\hline
    DARPA \cite{mit1999DARPA} & 1998/1999 & Open  & \cite{Schueller2018}&  Synthetic\\
    KDD-CUP'99 \cite{uci_kdd} & 1999 & Open & \cite{kabir_semiref}  & Synthetic \\
    Kyoto University \cite{kyoto} & Since 2009 & Open & \cite{Malaiya_semiref} & Real \\
    CTU-13 \cite{GARCIA2014100} & 2011 & Open & \cite{Timcenko2022,Chen2017} & Synthetic \\
    UNB ISCX IDS \cite{SHIRAVI2012357} & 2012 & Open & \cite{kabir_semiref} & Synthetic \\
    MAWI \cite{mawilab} & 2012 & Open &\cite{Kotani2018,GarciaCordero2016} & Real\\
    NSL-KDD \cite{LDhanabal2015ASO} & 2015 & Open & \cite{Li2024,Malaiya_semiref} & Synthetic \\
    UNSW-NB15 \cite{Moustafa2015} & 2015 & Open & \cite{Li2024,truonghuu_semiref} & Synthetic \\
    Stratosphere IPS \cite{stratodatasets} & 2015 & Open & \cite{truonghuu_semiref} & Synthetic \\
    UGR16 \cite{MACIAFERNANDEZ2018411} & 2016 & Open & \cite{Nguyen2019} & Real \\
    CIC-IDS-2017 \cite{Sharafaldin2018TowardGA} & 2017 & Open & \cite{Kompougias2021,Verkerken_semiref,Li2024,truonghuu_semiref} & Synthetic \\
    CIDDS-001 \cite{ring2017creation,ring2017flow}& 2017 & Open & \cite{GustavoGonz2021} & Synthetic\\
    CIC-DDoS2019 \cite{Sharafaldin} & 2019 & Open &  \cite{Li2024} & Synthetic\\
    ToN IoT \cite{Moustafa2019} & 2019 & Open & \cite{Gorbett2022} & Synthetic \\
    Aposemat IoT-23 \cite{garcia_dataset} & 2020 & Open & \cite{Gorbett2022} & Synthetic \\
    NF-BoT-IoT \cite{Sarhan} & 2021 & Open & \cite{Gorbett2022} & Synthetic \\
\hline
\end{tabular}
}
\caption{Results of the quality of primary articles}
\label{tab:datasets_discussion}
\end{table}


\section{Conclusions}
\label{sec:conclusions}

The utilization of machine learning, particularly through unsupervised learning methodologies, has emerged as a highly efficient technique for identifying malicious traffic within network flows. Furthermore, the availability of diverse datasets for model training and accuracy assessment underscores the robustness of this approach.

This article undertook a systematic review encompassing an initial pool of 63 articles, which, following rigorous evaluation processes and supplementary searches, culminated in the selection of 13 articles for detailed analysis.

In response to the first research question, 19 different machine learning algorithms have been identified that can be used to detect anomalous traffic. Autoencoders are the most commonly used solution to address this task, as 6 different articles have made use of them.

With respect to the datasets, 16 usable datasets for network anomaly detection have been found in the review answering the second research question. The data they contain, mostly synthetic, can be in the format of network flows or packets (pcap), in the last case, there are tools that take care of converting them into flows. In relation to the dates, datasets have been found between 1999 and 2021, all of which are free to download. Finally, the most widely used dataset is CIC-IDS-2017 as four articles have used it to test its accuracy. This is from 2017 and is made up of synthetic data.


\section*{Acknowledgments}


This publication is part of the CIBERLAB project, financed ``by European Union NextGeneration-EU, the Recovery Plan, Transformation and Resilience, through INCIBE''.  In addition, this work has been partially funded by the project EDMAR, PID2021-126592OB-C21, funded by MCIN/AEI/\\10.13039/501100011033 and by ERDF A way of making Europe.

%
%
%
\bibliographystyle{splncs03}
\bibliography{biblio}

\end{document}